\documentclass[12pt,a4paper]{article}
\usepackage{epsfig}
\begin{document}
\title{Non-universal gauge bosons $Z^{\prime}$ and rare top decays }

\author{Chongxing Yue $^{a}$, Hongjie Zong $^{b}$, Lanjun Liu $^{b}$\\
{\small $^{a}$ Department of Physics , Liaoning Normal University,
Dalian 116029, China
\thanks{E-mail:cxyue@lnnu.edu.cn}}\\
{\small $^{b}$ College of Physics and Information Engineering,}\\
\small{ Henan Normal University, Henan 453002, China } \\}

\date{\today}

\maketitle

\begin{abstract}
We study a new method for detecting non-universal gauge bosons
$Z'$ via considering their effects on rare top decays. We
calculate the contributions of the non-universal gauge bosons $Z'$
predicted by topcolor-assisted technicolor (TC2) models and
flavor-universal TC2 models on the rare top decays
$t\longrightarrow cV(V=g,\gamma,Z)$ and $t\longrightarrow
cl_{i}l_{j}(l_{i}, l_{j}=\tau, \mu$, or $e)$. We show that the
branching ratios of these processes can be significantly enhanced.
Over a sizable region of the parameter space, we have
$Br(t\longrightarrow cg)\sim10^{-5}$ and $Br(t\longrightarrow
c\tau\tau)\sim10^{-7}$, which may approach the observable
threshold of near future experiments. Non-universal gauge bosons
$Z'$ may be detected via the rare top decay processes at the
top-quark factories such as the CERN LHC.
\end{abstract}
\vspace{0.5cm} \noindent
 {\bf PACS number(s)}: 14.70.Pw, 12.60.Cn, 14.65.Ha

\newpage
\vspace{0.5cm}

~~ There are many models beyond the standard model (SM) predicting
the existence of extra $U(1)$ gauge bosons $Z'$. If discovered
they would represent irrefutable proof of new physics, most likely
that the SM gauge groups must be extended \cite{y1}. If these
extensions are associated with flavor symmetry breaking, the gauge
interactions will not be flavor-universal \cite{y2}, which predict
the existence of non-universal gauge bosons $Z'$. After the mass
diagonalization from the flavor eigenbasis into the mass
eigenbasis, the non-universal gauge interactions result in the
tree-level flavor-changing neutral current (FCNC) vertices of the
non-universal gauge bosons $Z'$. Hence, $Z'$ may have significant
contributions to some FCNC processes. If these effects are indeed
detected at LHC, LC or other experiments, it will be helpful to
identify the gauge bosons $Z'$, and therefore unravel underlying
theory \cite{y3}.

In the SM, due to the GIM mechanism, rare top decays induced by
FCNC are absent at tree-level and extremely small at loop levels.
In some new physics models beyond the SM, rare top decays may be
significantly enhanced \cite{y4}. Thus, rare top decays provide a
very sensitive probe of physics beyond the SM. Detection of rare top
decays at visible levels by any of the future colliders would be
instant evidence of new physics. Searching for rare top decays is
one of the major goals of the next generation of high energy
collider experiments, such as the Tevatron Run II and the CERN
LHC.

The rare top decays have been extensively studied in the context
of the SM and beyond \cite{y5}--\cite{y8}. It has been shown that
the SM predictions for rare top decays are far below the feasible
experimental possibilities at present or future collider
experiments and some models beyond the SM can enhance them by
several orders to make them approach the observable threshold of
near-future experiments. The aim of this paper is to show that the
non-universal gauge bosons $Z'$ predicted by topcolor-assisted
technicolor (TC2) models \cite{y9} and flavor-universal TC2 models
\cite{y10} can give significant contributions to the rare top
decays $t\longrightarrow cV (V=g,\gamma$  or $Z)$ and
$t\longrightarrow cl_{i}l_{j} (l_{i}, l_{j}=\tau, \mu$, or $e)$.
With reasonable values of the parameters, the value of
$Br(t\longrightarrow cg)$ can reach $5.1\times10^{-5}$. For the
branching ratio $Br(t\longrightarrow c\tau\tau)$, it is possible
to reach $6.4\times10^{-6}$. Thus, rare top decays may be used to
detect non-universal gauge bosons $Z'$ in near future experiments.

A common feature of TC2 models \cite{y9} and flavor-universal TC2
models \cite{y10} is that a large part of the top quark mass is
dynamically generated by topcolor interactions at a scale of order
$1TeV$. To ensure that the top quark condenses and receives a
large mass while the bottom quark does not, a non-universal
extended hypercharge gauge group $U(1)$ is often invoked, so that
the topcolor gauge group is usually taken to be a strong coupled
$SU(3)\times U(1)$. Breaking of the extended gauge groups to their
diagonal subgroups produces non-universal massive gauge bosons.
Thus, TC2 models and flavor-universal TC2 models all predict the
existence of the non-universal gauge bosons $Z'$. These new
particles treat the third generation fermions differently from
those in the first and second generations. So, they can lead to FC
couplings.

The flavor-diagonal couplings of $Z'$ to ordinary fermions, which
are related to our calculations, can be written as \cite{y11}:
\begin{eqnarray}
{\cal
L}^{FD}_{Z'}&=&g_{1}cot\theta'Z'_{\mu}[(\frac{1}{6}\bar{t}_{L}\gamma^{\mu}t_{L}+
\frac{2}{3}\bar{t}_{R}\gamma^{\mu}t_{R})\nonumber\\
&&-(\frac{1}{2}\bar{\tau}_{L}\gamma^{\mu}\tau _{L}+
\bar{\tau}_{R}\gamma^{\mu}\tau_{R})\nonumber\\
&&-tan^{2}\theta'(\frac{1}{2}\bar{\mu}_{L}\gamma^{\mu}\mu_{L}
+\bar{\mu}_{R}\gamma^{\mu}\mu_{R})\nonumber\\
&&-tan^{2}\theta'(\frac{1}{2}\bar{e}_{L}\gamma^ {\mu}e_{L}
+\bar{e}_{R}\gamma^{\mu}e_{ R})],
\end{eqnarray}
where $g_{1}$ is the hypercharge gauge coupling constant,
$\theta'$ is the mixing angle with $tan\theta'=g_{1}/\sqrt{4\pi
k_{1}}$. To obtain the top quark condensation and not form a
$b\bar{b}$ condensation, there must be $tan\theta'<<1$
\cite{y9,y10}. The FC couplings of $Z'$ to ordinary fermions, which
are related to  rare top decays, can be written as:
\begin{eqnarray}
{\cal
L}^{FC}_{Z'}&=&-\frac{1}{6}g_{1}K_{tc}Z'_{\mu}(\bar{t}_{L}\gamma^{\mu}c_{L}
+4\bar{t}_{R}\gamma^{\mu}c_{R})\nonumber\\
&&-\frac{1}{2}g_{1}Z'_{\mu}[K_{\tau\mu}
(\bar{\tau}_{L}\gamma^{\mu}\mu_{L}+2\bar{\tau}_{R}\gamma^{\mu}\tau_{R})\nonumber\\
&&+K_{\tau
e}(\bar{\tau}_{L}\gamma^{\mu}e_{L}+2\bar{\tau}_{R}\gamma^{\mu}e_{R})\nonumber\\
&&+tan^{2}\theta'K_{\mu
e}(\bar{\mu}_{L}\gamma^{\mu}e_{L}+2\bar{\mu}_{R}\gamma^{\mu}e_{R})],
\end{eqnarray}
where $K_{tc}$ and $K_{ij}$ are the flavor mixing factors.

From Eq.(1) and Eq.(2) we can see that the rare top decays
$t\longrightarrow cV$ and $t\longrightarrow cl_{i}l_{j}$ can be
generated via $Z'$ exchange in TC2 models and flavor-universal TC2
models. We first consider the rare top decay processes
$t\longrightarrow cV (V=g, \gamma$, or $Z)$. These processes are
induced by penguin diagrams. The calculations are straightforward
but tedious. Similarly Feynman diagrams for contributions of the
neutral scalars to $t\longrightarrow cV$  and $t\longrightarrow
ch$ have been calculated in Ref. \cite{y6}. Here we do not present
the lengthy formulas which are expressed in terms of two- and
three- point standard Feynman integrals.

\begin{figure}[h]
\begin{center}
\vspace*{-0.5cm} \epsfig{file=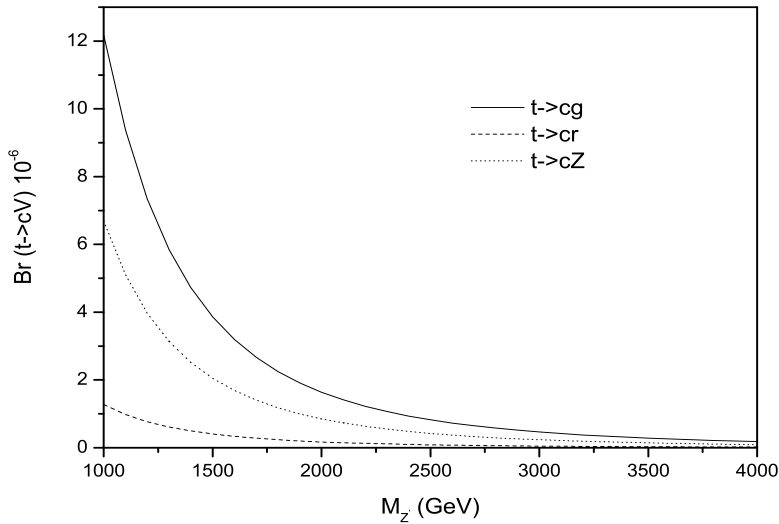,width=330pt,height=230pt}
\vspace*{-1.2cm} \caption{Branching ratios $Br(t\longrightarrow
cV)$ as functions of the gauge bosons $Z'$ mass $M_{Z'}$ for the
flavor mixing factor $K_{tc}=\lambda=0.22$.} \label{zt1.eps}
\end{center}
\end{figure}

\begin{figure}[h]
\vspace*{-1.5cm}
\begin{center}
\epsfig{file=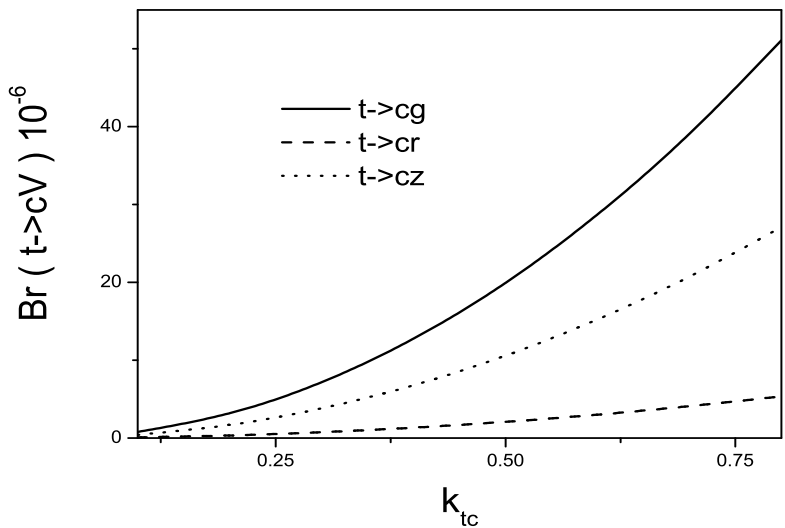,width=330pt,height=230pt}
 \vspace*{-1.2cm}
\caption{Branching ratios $Br(t\longrightarrow cV)$
        as functions of  $k_{tc}$ for $M_{Z'}=1.5TeV$.}
 \label{zt2}
\end{center}
\end{figure}
 \vspace*{-1cm}

Our numerical results are summarized in Fig.1 and Fig.2. In Fig.1,
we have assumed $K_{tc}=\lambda$, which $\lambda=0.22$ is the
Wolfenstein parameter \cite{y12}. The limits on the mass of the
extra $U(1)$ gauge bosons $Z'$ can be obtained via studying their
effects on various experimental observables \cite{y11}. For
example, Ref. \cite{y2} studied the bound placed by the
electroweak measurement data on the $Z'$ mass $M_{Z'}$. They find
that $Z'$ predicted by TC2 models and flavor-universal TC2 models
must be heavier than about $1TeV$. As estimation the contributions
of $Z'$ to the rare top decays, we have taken the $Z'$ mass
$M_{Z'}$ as a free parameter in Fig.1 and Fig.2. One can see from
Fig.1 that the branching ratio is sensitive to the $Z'$ mass
$M_{Z'}$ and suppressed by large $M_{Z'}$. For $1TeV\leq
M_{Z'}\leq2TeV$, the values of the branching ratios
$Br(t\longrightarrow cg)$, $Br(t\longrightarrow c\gamma)$, and
$Br(t\longrightarrow cZ)$ vary in the ranges of
$1.6\times10^{-6}\sim1.2\times10^{-5}$,
$1.7\times10^{-7}\sim1.3\times10^{-6}$, and
$8.5\times10^{-7}\sim6.7\times10^{-6}$, respectively. To see the
effects of the flavor mixing factor $K_{tc}$ on the branching
ratios $Br(t\longrightarrow cV)$,  we plot $Br(t\longrightarrow
cV)$ as functions of $K_{tc}$ for $M_{Z'}=1.5TeV$ in Fig.2. One
can see from Fig.2 that the values of $Br(t\longrightarrow cg)
[Br(t\longrightarrow c\gamma )]$ can reach $5.1\times10^{-5}$
$[5.3\times10^{-6}]$ for $M_{Z'}=1.5TeV$ and $K_{tc}=0.8$.

The LHC sensitivity to the top quark anomalous couplings
$K_{tq}^{g}$ and $K_{tq}^{\gamma}$, which come from the effects of
new physics, can be obtained by studying quark-gluon fusion,
single top quark production, $t+\gamma(Z)$ production, and
like-sign top-pair final states, separately. The relevant results
are summarized in Ref. \cite{y13}. If we assume that the anomalous
coupling vertexes $\bar{t}cV$ come from the non-universal gauge
bosons $Z'$, the contributions of $Z'$ to the branching ratios
$Br(t\longrightarrow cV)$ can be transposed to those of $Z'$ to
the top quark anomalous couplings $K_{tq}^{g}$ and
$K_{tq}^{\gamma}$. For example, the value of the anomalous
coupling $K_{tc}^{g}$ can reach $6.8\times10^{-3}$ for
$M_{Z'}=1.5TeV$ and $K_{tc}=0.8$. Hence, we can conclude that, if
non-universal gauge bosons $Z'$ are not too heavy and can induce
large FC couplings, the possible signals of these new particles
may be detected via the FCNC processes $t\longrightarrow cV$ at
the top quark factories such as  the CERN LHC.

 The non-universal gauge bosons $Z'$ can induce
the tree-level FC couplings at quark and lepton sector. Thus, the
rare top decays $t\longrightarrow cl_{i}l_{j}$ can be generated at
tree-level or at one-loop only for $i=j$. However, the
contributions of $Z'$ to the processes $t\longrightarrow
cl_{i}l_{j}$ via the off-shell photon pengiun and $Z$ pengiun
diagrams are only at the order of the magnitude of $1\%$ of the
tree-level result and therefore can be ignored. The conclusion is
similar to that of the lepton flavor-violation tau decays
$\tau\longrightarrow l_{i}l_{j}l_{k}$ \cite{y14}.

In the context of TC2 models and flavor-universal TC2 models, the
non-universal gauge bosons $Z'$ only treat the fermions in the
third generation differently from those in the first and second
generation and treat the fermions in the first generation same as
those in the second generation. Thus, there must be
$Br(t\longrightarrow c\mu\mu)=Br(t\longrightarrow cee)$ and
$Br(t\longrightarrow c\tau\mu)=Br(t\longrightarrow c\tau e)$ for
assuming $K_{\tau\mu}=K_{\tau e}$. Compared with the couplings of
$Z'$ to the third family fermions, the couplings of $Z'$ to the
first and second family fermions are suppressed by the factor
$tan^{2}\theta'$. It has been shown that the choice $k_{1}=1$
corresponds to $tan^{2}\theta'\approx0.01$ \cite{y11}. Thus, the
branching ratio $Br(t\longrightarrow c\tau\tau)$ is about four
orders of magnitude larger than those of the processes
$t\longrightarrow c\mu\mu$ and $t\longrightarrow cee$.
\begin{figure}
\vspace*{0cm}
\begin{center}
 \epsfig{file=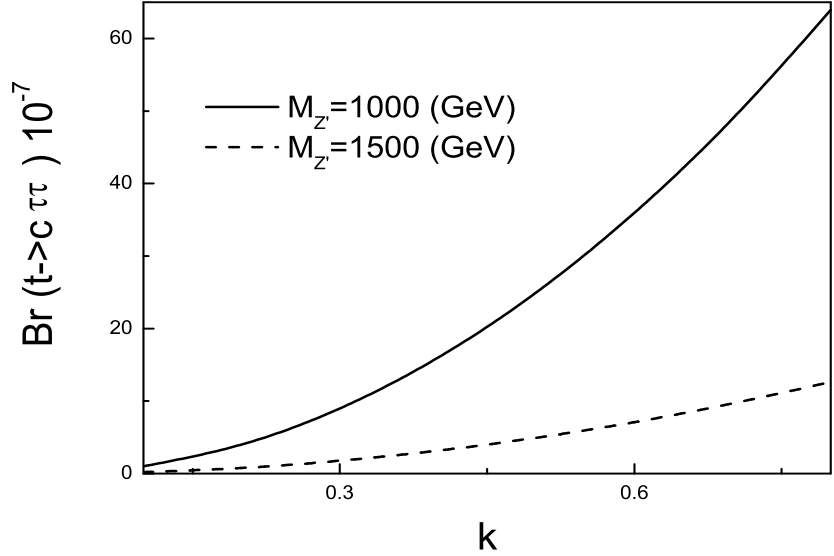,width=350pt,height=250pt}
\vspace*{-1.2cm} \caption{Branching ratio of the process
$t\rightarrow
        c\tau\tau$ as a function of the flavor mixing $K$
        for $M_{Z'}=1.0TeV$(solid line) and  $1.5TeV$(dashed line).}
\label{zt3}
\end{center}
 \vspace*{-0.5cm}
\end{figure}

\vspace*{0cm}

 In Fig.3 and Fig.4 we plot branching ratios
$Br(t\longrightarrow c\tau\tau)$ and $Br(t\longrightarrow
c\tau\mu)$ as functions of the flavor mixing factor $K$. We have
assumed in Fig.3 and Fig.4 that the flavor mixing factor in quark
sector is equal to that in lepton sector i.e.
$K_{tc}=K_{\tau\mu}=K$. One can see that the value of
$Br(t\longrightarrow c\tau\tau)$ is larger than that of
$Br(t\longrightarrow c\tau\mu)$ in all of the parameter space. For
$M_{Z'}=1.5TeV$ and $0.1\leq K\leq\frac{1}{\sqrt{2}}$, there are
$2\times10^{-8}\leq Br(t\longrightarrow
c\tau\tau)\leq9.7\times10^{-7}$ and $2\times10^{-12}\leq
Br(t\longrightarrow c\tau\mu)\leq4.8\times10^{-9}$. For
$M_{Z'}=1.0TeV$ and $K=0.8$, it is possible to have
$Br(t\longrightarrow c\tau\tau) =6.4\times10^{-6}$ and
$Br(t\longrightarrow c\tau\mu)=4.2\times10^{-8}$. Certainly, the
values of branching ratios $Br(t\longrightarrow cl_{i}l_{j})$ are
strongly dependent on the value of the flavor mixing factor $K$.
We expect that this physical parameter can be measured in the
future experiments and it can give a strong clue about the new
physics beyond the SM.

\begin{figure}
\begin{center}
\epsfig{file=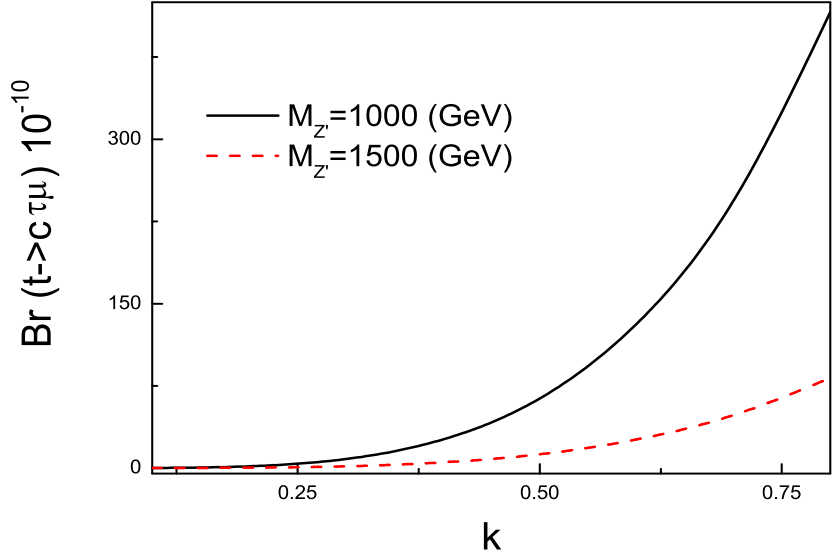,width=350pt,height=250pt} \vspace*{-1.2cm}
\caption{Same as Fig.3 but for $Br(t\longrightarrow c\tau\mu)$.}
\label{zt4}
\end{center}
 \vspace*{-0.5cm}
\end{figure}

If the top anomalous couplings are present and large enough, the
new particles  can give significantly contributions to single
production of top quarks. These have been extensively studied at
lepton colliders \cite{y15} and at hadron colliders \cite{y16}.
Thus the non-universal gauge bosons $Z'$ predicted by TC2 models
and flavor-universal TC2 models might give large contributions to
single production of top quarks.

  To summarize, we have studied the contributions of the non-universal
gauge bosons $Z'$ predicted by TC2 models and flavor-universal TC2
models to the rare top decays $t\longrightarrow cV$ and
$t\longrightarrow cl_{i}l_{j}$. We have shown that the branching
ratios of these processes can be significantly enhanced. With
reasonable values of the parameters, the branching ratio
$Br(t\longrightarrow cg)$ $[Br(t\longrightarrow c\tau\tau)]$ can
reach $5.1\times10^{-5}$ $[6.4\times10^{-6}]$, which might
approach the observable threshold of near future experiments. For
the branching ratio $Br(t\longrightarrow c\tau\mu)$, it is
possible to reach $4.1\times10^{-8}$. The production of
$10^{7}\sim10^{8}$ top quark pairs per year at the CERN LHC will
allow to probe the top couplings to both known and new particles
involved in possible top decay channels different from the main
$t\longrightarrow Wb$. On a purely statistical basis, one should
be able to detect a particular decay channel whenever its
branching ratio is larger than about $10^{-6}\sim10^{-7}$
\cite{y4,y17}. Hence, the non-universal gauge bosons $Z'$ may be
detected via rare top decays at the top-quark factories, the CERN
LHC, as long as their masses are not too heavy.

Except TC2 models and flavor-universal TC2 models, there are many
models beyond the SM, such as non-commuting ETC models\cite{y18}
and top flavor seesaw models\cite{y19}, predicting the existence
of non-universal gauge bosons $Z'$, which have similar property
considered in this paper. Therefore, we believe that our
conclusion is quite model-independent.

\vspace{1.0cm} \noindent{\bf Acknowledgments}

This work was supported in part by National Natural Science Foundation
of China (90203005).


\vspace*{2cm}
\begin{thebibliography}{99}
\bibitem{y1}J. Erler and P. Langacker, {\em Phys. Lett. B}{\bf 456}, 68(1999);
            {\em Phys.  Rev.   Lett.}{\bf 84}, 212(2000);
            M. Chanowitz, {\em  Phys. Rev. Lett.}{\bf 87}, 231802(2001).
\bibitem{y2}R. S. Chivukula and E. H. Simmons, {\em Phys. Rev. D}
            {\bf 66}, 015006(2002); R.S. Chivukula, H.-J. He,
            J. Howard, E.H. Simmons, hep-ph/0307209.
\bibitem{y3}A. A. Pankov, hep-ph/9910507; S. Godfrey, hep-ph/0201092.
\bibitem{y4}For review see B. Mele, hep-ph/0003064; S. Bejar, J. Guasch,
            and J. Sola, hep-ph/0111294.
\bibitem{y5}G. Eilam, J. L. Hewett and A. Soni, {\em Phys. Rev.
            D}{\bf 44}, 1473(1991); E. Jenkins, {\em Phys. Rev. D}
            {\bf 56}, 458(1997); B. Mele, S. Petrarca, A.
            Soddu, {\em Phys. Lett. B}{\bf 435}, 401(1998).
\bibitem{y6}W. S. Hou, {\em Phys. Lett. B}{\bf 296}, 179(1992);
            M. Luke and M. J. Savage, {\em Phys. Lett. B}{\bf 307}, 387(1993);
            D. Atwood, L. Reina, and A. Soni, {\em Phys. Rev. D}{\bf
            55}, 3156(1997); Chongxing Yue, Gongru Lu, Gouli
            Liu, and Qingjun Xu, {\em Phys. Rev. D}{\bf 64},
            095004(2001); E. Iltan, {\em Phys. Rev. D}{\bf
            65}, 075017(2002); E. O. Iltan and I. Turan,
            {\em Phys. Rev. D}{\bf67}, 015004(2003).
\bibitem{y7}C. S. Li, R. J. Oakes and J. M. Yang, {\em Phys. Rev.
            D}{\bf 49}, 293(1994); G. Couture, C. Hamzaoui and H.
            K$\ddot{o}$nig, {\em Phys. Rev. D}{\bf 52}, 171(1995);
            J. L. Lopez, D. V. Nanopoulos and R. Rangarajan, {\em Phys. Rev.
            D}{\bf 56}, 3100(1997); G. Couture, M. Frank and H.
            K$\ddot{o}$nig, {\em Phys. Rev. D}{\bf 56}, 4213(1997);
            G. M. de Divitiis, R. Petronzio, L. Silvestrini, {\em Nucl. Phys.
            B}{\bf 504}, 45(1997); J. M. Yang, B.-L. young and X.
            Zhang, {\em Phys. Rev. D}{\bf 58}, 055001(1998);
            T. Han and Jing Jiang, B. Mele, {\em Phys. Lett. B}{\bf 516}, 337(2001).
\bibitem{y8}T. Han, R. D. Peccei, and X. Zhang, {\em Nucl. Phys.
            B}{\bf 454}, 527(1995); B.-L. Young, and X. Zhang, {\em Phys. Rev.
            D}{\bf 55}, 7241(1997); M. Hosch, K. Whisnant, and B.-L. Young,
            {\em Phys. Rev. D}{\bf 56}, 5725(1997);
            E. Malkawi and T. Tait, {\em Phys. Rev. D}{\bf 54},
            5758(1996); T. Tait and C. P. Yuan, {\em Phys. Rev. D}{\bf
            55}, 7300(1997); T. Han, M. Hosch, K. Whisnant, B.-L.
            young and X. Zhang, {\em Phys. Rev. D}{\bf 58},
            073008(1998); F. del Aguila and J. A.
            Aguilar-Saavedra, {\em Phys. Lett. B}{\bf 462},
            310(1999); J. L. Diaz-Cruz, M. A. Perez, G.
            Tavares-Velasco, J. J. Toscano, {\em Phys. Rev. D}{\bf 60},
            115014(1999); F. del Aguila and J. A.
            Aguilar-Saavedra, {\em Nucl. Phys. B}{\bf
            576}, 56(2000); Chongxing Yue, Guoli Liu, Qingjun Xu, {\em Phys. Lett. B}{\bf
            508}, 290(2001).
\bibitem{y9}C. T. Hill, {\em Phys. Lett. B}{\bf 345}, 483(1995);
            K. Lane and E. Eichten, {\em Phys. Lett. B}{\bf 352}, 383(1995);
            K. Lane, {\em Phys. Lett. B}{\bf 433}, 96(1998) G. Cvetic,
            {\em Rev. Mod. Phys.} {\bf 71}, 513(1999).
\bibitem{y10}M. B. Popovic and E. H. Simmons, {\em Phys. Rev. D}{\bf
            58}, 095007(1998); G. Burdman and N. Evans, {\em Phys. Rev. D}{\bf
            59}, 115005(1999).
\bibitem{y11}G. Buchalla, G. Burdman, C. T. Hill, D. Kominis, {\em Phys. Rev. D}
            {\bf 53}, 5185(1996); C. T. Hill and E. H. Simmons, {\em Phys.
            Rep.}{\bf381}, 235(2003).
\bibitem{y12}L. Wolfenstein, {\em Phys. Rev. Lett.}{\bf
             51}, 1945(1983).
\bibitem{y13}M. Beneke et al. and A. Ahmadov et al., "Top Quark Physics",
             hep-ph/0003033.
\bibitem{y14}Chongxing Yue, Yanming Zhang, Lanjun Liu, {\em Phys. Lett.
             B}{\bf 547}, 252(2002).
\bibitem{y15}V. F. Obraztsov,et al., {\em Phys. Lett. B}{\bf 426}, 393(1998);
             T. Han and J. L. Hewett, {\em Phys. Rev. D}{\bf 60}, 074015(1999);
             A. T. Alan and A. Senol, {\em Europhys. Lett.}{\bf 57}, 669(2001).
\bibitem{y16}T. Tait, C. P. Yuan, {\em Phys. Rev. D}{\bf 55}, 7300(1997);
             T. Tait, C. P. Yuan, {\em Phys. Rev. D}{\bf 63}, 014018(2001).
\bibitem{y17}Tao Han and D. Marfatia, {\em Phys. Rev. Lett.}{\bf
             86}, 1442(2001); Tao Han and B. McElrath, {\em Phys. Lett.
             B}{\bf 528}, 81(2002).
\bibitem{y18}R. S. Chivukula, E. H. Simmons, and J. Terning, {\em Phys. Lett. B}
             {\bf 331}, 383(1994).
\bibitem{y19}H. J. He, T. M. P. Tait, C. P. Yuan, {\em Phys. Rev. D}{\bf 62}, 011702(2000).
\end{thebibliography}
\end{document}